\documentclass[english,twocolumn,showpacs,prb,amsmath,amssymb]{revtex4}
\usepackage[latin9]{inputenc}
\usepackage{graphicx}
\usepackage{amssymb}
\usepackage{epsfig}
\usepackage{dcolumn}

\makeatletter

\begin{document}

\title{Magnetism and magnetotransport in disordered graphene}

\author{T. G. Rappoport$^{1}$, Bruno Uchoa$^{2}$, and A. H. Castro Neto$^{3}$}

\affiliation{$^{1}$ Instituto de F\'{\i}sica, Universidade Federal do Rio de
Janeiro, Rio de Janeiro, RJ, 68.528-970, Brazil \\
 $^{2}$Department of Physics, University of Illinois at Urbana-Champaign,
1110 W. Green St., Urbana, IL 61801-3080, USA \\
 $^{3}$Department of Physics, Boston University, 590 Commonwealth
Avenue, Boston, MA 02215, USA}

\date{\today}

\begin{abstract}
We perform Monte Carlo simulations to study the interplay of structural
and magnetic order in single layer graphene covered with magnetic
adatoms. We propose that the presence of ripples in the graphene structure
can lead to clustering of the adatoms and to a variety of magnetic
states such as super-paramagnetism, antiferromagnetism, ferromagnetism
and spin glass behavior. We derive the magnetization hysteresis and also
the magnetoresistance curves in the variable range hopping regime,
which can provide experimental signatures for ripple induced clustering
and magnetism. We propose that the magnetic states in graphene can
be controlled by gate voltage and coverage fraction. 
\end{abstract}


\maketitle

\section{Introduction}

Graphene is probably one of the most remarkable discoveries in condensed
matter physics in the last decade \cite{geim_review,rmp}. The material
is a two-dimensional (2D) crystal composed of carbon (C) atoms with
sp$^{2}$ hybridization, that is, graphene is one atom thick and thus
the thinnest cloth in nature. Because of its low dimensionality, it
does not show structural long-range order in its free form, as per
the Hohenberg-Mermin-Wagner (HMW) theorem\cite{HMM}, but it can
present a flat phase at low temperatures due to non-linear effects
or in the presence of a substrate, scaffolds, contacts, or impurities
that break explicitly the translational symmetry perpendicular to
the graphene plane\cite{chaikin}. The leftovers of the fluctuations
that forbid long-range order are found in the form of frozen ripples
in suspended \cite{meyer07,philip} samples.

While it was theoretically predicted early on that graphene by itself
would not be magnetic\cite{peres_ferro}, it has been shown that
adatoms in the graphene surface can easily form magnetic moments due
to graphene's unusual electronic properties such as low density of
states and chirality\cite{anderson_graphene}. Moreover, because
graphene has a low density of states close to the Dirac point, the
Kondo effect is suppressed\cite{fradkin,graphene_kondo}, allowing for the
appearance of magnetic states. On the other hand, because of its low
dimensionality, long-range magnetic order is inhibited, and the intrinsic
coupling between structural and magnetic fluctuations may lead to
very inhomogeneous spin textures (such as of Griffiths phases~\cite{griffiths})
that can be found in complex itinerant magnetic systems such as disordered
Kondo Lattices\cite{griffiths-ah}.

In this paper, performing a series of Monte Carlo simulations over
disordered realizations of magnetic adatoms in a graphene sheet, we
examine the magnetic correlations of the local moments in the magnetization
and in the magnetoresistance curves. In the case of adatoms such as
Hydrogen (H), where the probability of adsorption changes substantially
according to the local curvature of the graphene sheet, we show that
the formation of local magnetic moments leads to an interesting interplay
between the correlation due to the RKKY interaction and the ripples,
generating a variety of magnetic textures, tendency to clustering
at low concentrations and a percolative transition at higher concentration
of adatoms. Since the overall magnetic response of a single layer
is rather small compared to the usual response in bulk, magnetotransport
measurements are probably the easiest way to probe magnetic correlations
in graphene. Based in our Monte Carlo results for the magnetization,
we calculate the magnetoresistance curves in graphene for strong disorder.
These curves can offer clear experimental signatures for the presence
of macroscopic magnetic states in graphene in the regime of variable
range hopping~\cite{elias09}.

The structure of the paper is as follows: in sec. II we introduce
the spin Hamiltonian for the magnetic adatoms; in sec. III we derive
the heuristic rules for adsorption of adatoms in graphene, comparing
different realizations of disorder for different adsorption probability
distributions, which we define in terms of the curvature (height)
of the ripples. In sec. IV, we derive the magnetization curves and
in sec. V we calculate the magnetoresistance. Finally, in sec. VI
we present our conclusions.

\section{RKKY Hamiltonian}

Our starting point is the tight-binding Hamiltonian of the electrons
in graphene (we set $\hbar=1$)\cite{rmp}: \begin{eqnarray}
{\cal H}_{TB} & = & -t\sum_{\sigma}\sum_{\langle i,j\rangle}\left[a_{\sigma}^{\dagger}(\mathbf{R}_{i})b_{\sigma}(\mathbf{R}_{j})+{\rm h.c.}\right],\label{HTB}\end{eqnarray}
 where $a_{\sigma}({\bf R}_{i})$ ($b_{\sigma}({\bf R}_{i})$) annihilates
and electron with spin $\sigma=\uparrow,\downarrow$ on sublattice
$A$ ($B$) at position $\mathbf{R}_{i}$, and $t$ ($\approx2.7$
eV) is the nearest neighbor hopping energy. When adatoms are added
to graphene they can hybridize with an energy $V$ to the C atoms
and if the local Coulomb energy $U$ is sufficiently large, a local
moment of spin $S$ is formed\cite{anderson_graphene} at a site
${\bf R}_{i}$. This spin $S_{i}$ interacts with the graphene electrons
via an exchange interaction, $J_{k}\approx-V^{2}/U$, which is described
by the Hamiltonian: \begin{eqnarray}
{\cal H}_{s}=J_{k}\sum_{i}S_{i}\cdot s_{i}\,,\label{kondo}\end{eqnarray}
 where $s_{i}$ is the graphene electron spin. Eq. (\ref{HTB}) together
with (\ref{kondo}) describe a Kondo lattice in graphene.

The Kondo interaction (\ref{kondo}) induces an indirect kinetic exchange
interaction between local moments, the RKKY interaction, which depends
on the chemical potential $\mu$ and hence can be controlled with
an external gate voltage. Therefore, the nature of the magnetic states
in graphene can be controlled by the application of a transverse \textit{electric}
field, a situation that never occurs in metals. For low carrier concentrations,
i.e. close to the Dirac point, the interaction between spins located
in the same sublattice is ferromagnetic, $\chi_{AA}(\mathbf{R})=\chi_{BB}(\mathbf{R})\propto1/R^{3}$,
while it is antiferromagnetic if they belong to opposite sublattices
($\chi_{AB}(\mathbf{R})\propto-1/R^{3}$)\cite{brey,falko,saremi,costa}.
Given the position of the magnetic moments and the effective interaction
between them, we can perform numerical simulations to obtain the magnetic
properties of the system.

When an adatom sits on top of a carbon atom, the carbon sp$^{2}$bonds
are locally distorted and acquire a sp$^{3}$ character as in diamond,
with the carbon underneath the impurity being pulled out of the plane\cite{hallmark}.
Although a flat graphene sheet has almost no magnetic anisotropy due
to the very small spin-orbit coupling\cite{macdonald}, in the presence
of ripples\cite{paco}, lattice distortions of the sp$^{3}$ type
that are generated by the presence of adatoms such as H can substantially
increase the spin-orbit interaction close to the adatoms, generating
a local spin-orbit coupling up to $\Delta_{SO}\approx7$ meV\cite{sp3so},
which corresponds to an out of plane magnetic anisotropy of $\approx7$
T. In the range of temperature $k_{B}T\lesssim\Delta_{SO}$, the minimal
model for the interaction between adatoms in graphene should be correctly
captured by the physics of the Ising model: \begin{equation}
{\cal H}_{{\rm eff}}=\sum_{i,j}J_{{\rm RKKY}}(r_{ij})S_{i}^{z}S_{j}^{z}-g\mu_{B}H_{ex}\sum_{i}S_{i}^{z}.\end{equation}
 where $J_{{\rm RKKY}}(r_{ij})$ is a spatially dependent RKKY interaction
that depends on $\mu$ and $H_{ex}$ is an external magnetic field.
For $\mu\ll t$, the period of the RKKY oscillation is long compared
to the typical atomic distances and we only need to consider the power-law
decay of the interaction. The exchange is given by $J_{AA}(r)=J_{BB}(r)=J_{0}\exp(-r_{ij}/r_{0})/r_{ij}^{3}$
for spins in the same sublattice and $J_{AB}(r)=-J_{0}\exp(-r_{ij}/r_{0})/r_{ij}^{3}$
for spins in different sublattices. $r_{ij}$ is given in units of
the lattice constant $a=2.46$ \AA\ and $r_{0}\approx12$ \AA \ is
a cut-off distance that is introduced to limit the range of the parameters
in the numerical calculations.

\section{Ripples and adsorption probability}

In a flat graphene layer, the magnetic interaction between impurity
spins would generate uniform magnetic states. Nevertheless, ripples
in the graphene structure break the translational symmetry and lead
to a inhomogeneous situation where the adatoms have preferential sites
for hybridization. Due to ripples, the topography of the surface in
graphene is not flat but \textit{\emph{has}} \textit{curvature,} \textit{\emph{which
appears in}} in the form of {}``valleys'' and {}``hills''. In
graphene, the typical ratio between the height ($h$) and the lateral
size ($L$) of the ripples is $h/L\approx0.1-0.2$ \cite{meyer07}.
On top of a hill, the characteristic curvature of the ripples distorts
the sp$^{2}$ bonds to an extent that they acquire sp$^{3}$ character,
with a significant hybridization between the $\sigma$ and $\pi$
bands. In this situation, the adsorption of a H atom on a hill-C has
been shown to be energetically favorable, helping to stabilize the
ripple\cite{curve_graphene}. On the other hand, for perfectly flat
graphene, the adsorption of a H atom costs some energy to locally
distort the sp$^{2}$ bonds and pull the hydrogenated C atom out of
the plane. The difference in the chemisorption energy between the
two cases, recently calculated by \textit{ab initio} methods, can
be as large as 2.5 eV. These results also indicate the existence of
a minimal curvature for the ripples ($h/L\gtrsim0.12$) above which
the adsorption of a H atom can be strongly favorable\cite{curve_graphene}.
The substantial change in the chemisorption energy of H with the local
curvature in graphene is qualitatively consistent with previous theoretical
~\cite{curve_nano1} and experimental \cite{curve_nano2} results
in nanotubes, where the binding energy has been shown to change dramatically
with the radius of the tube.

Assuming that graphene is supported on a substrate, only one surface
is available for hybridization. Notice that because of curvature,
the $p_{z}$ orbitals that are locally perpendicular to the graphene
surface, approach each other in the valleys but distance themselves
in the hills. Contrary to the hill-C case, the local curvature is
expected to inhibit the H adsorption in the valleys, since in order
to stabilize the C-H bond relative to the other three C sp$^{3}$-like
bonds, the H has to pull the C atom out of the plane against the curvature,
at the expense of an additional cost of energy, compared to the flat
case (the difference in chemisorption energies between the hill-C
and the valley-C cases could easily amount to several eV). The H adatoms
therefore hybridize much more easily in the hills than in the valleys,
leading to a {}``percolative'' structure as the planar density of
adatoms is increased. Therefore, the probability that an adatom hybridizes
with a hill-C is larger than a valley-C atom.

\begin{figure}[t]
 \vspace{-0.2cm}
 \includegraphics[clip,width=0.95\columnwidth]{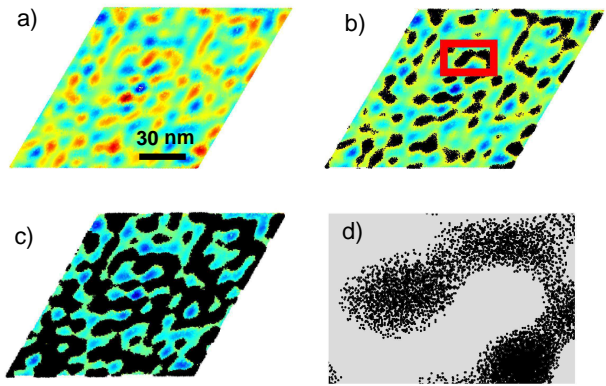}

\caption{(a) Graphene sheet with random ripples. The color map represents the
height increasing from blue to red in the interval $z\in[-z_{max},z_{max}]$.
(b) -(c) graphene sheet covered with adatoms for a lower cut-off $h_{0}=0.28\, z_{max}$
and $h_{0}=0$ respectively (see main text). (d) Zoom-in showing the
adatoms on top of a hill (red rectangle).\label{fig1}}

\end{figure}

Let us assume, for the sake of the argument, that the ripples in graphene
have a simple sinusoidal form (see ref. \onlinecite{meyer07}),
that is, if $z$ is the perpendicular coordinate of the C atom relative
to the flat situation, then $z(x,y)=A\sin(k_{x}x+k_{y}y+\phi)$ where
$A$, $\phi$, $k_{x}$ and $k_{y}$ are random parameters. A large
number (up to 200) of such sinusoidal waves can be superimposed to
obtain a randomly curved sheet, as shown in Fig. ~\ref{fig1} (a)
for a system with $2\times600\times600$ C atoms. Next, we decide
a strategy to allow the incorporation of adatoms by the graphene sheet
for some quenched realization of ripples.

Given the existence of a lower bounded range of local curvature where
the adsorption is most likely to happen\cite{curve_graphene}, and
also the fact that the adsorption is much easier on the top of a hill
rather than in the bottom of a valley in graphene, we assume a probability
distribution that grows monotonically with the height of the ripples
and has some effective lower cut-off, $h_{0}$, below which the adsorption
is unlikely to happen. As a toy model, we may assume for instance
that the probability of adsorption varies linearly with $z$ through
the following criterion: we allow the incorporation if \begin{equation}
z>h_{0}+h_{i},\end{equation}
 where $h_{i}$ is a random number varying from $0$ to $z_{max}-h_{0}$,
and $h_{0}\in[-z_{max},z_{max}]$ is the lower cut-off of the distribution,
with $z_{max}$ as the maximum height. The choice of another distribution
function such as, for instance, a step function, $\theta(z-h_{0})$,
or some other distribution with a finite tail for $z<h_{0}$, should
lead to similar conclusions regarding the physical properties, as
far as the probability of adsorption on the top of the hills is large
compared to the probability adsorption in the valleys.

If the number of atoms which are available for adsorption is fixed,
by varying $h_{0}$ we obtain different incorporation fractions, as
shown in Figures~\ref{fig1} (b),(c). Notice that for a given realization
of ripples, as one increases the adatom concentration, one obtains
a percolative structure. At low coverage densities the adatoms form
\textit{clusters}. Hence, ripples naturally lead to clustering. We
can also fix the total number of adatoms attached to the sample and
see how the different probability functions change the coverage structure
of the system. In this case, the overall structure is the same of
Figures~\ref{fig1} (b),(c) but different probability distributions
lead to different concentration of adatoms on top of the hills, as
it will be clear in the next section.

\section{Magnetic textures}

The details of the clustering depend on the nature of the adatoms
used. For instance, if one uses H atoms, one might imagine that only
H atoms that are separated by more than one lattice spacing are stable
since H atoms which are in neighboring C sites can recombine into
H$_{2}$ molecules and leave the graphene surface~\cite{h2}. Hence,
we have considered two different situations, namely, either the adatoms
have nearest neighbors or not. We studied the location of magnetic
adatoms using the same incorporation rules discussed above, with and
without allowing the existence of nearest neighbors. For the system
without first neighbors, we randomly remove the neighbors, in order
to avoid any artificial distribution of spins. We found that the magnetic
response of these two systems are rather different and can be distinguished
experimentally.

\begin{figure}[b]
\includegraphics[clip,width=0.8\columnwidth]{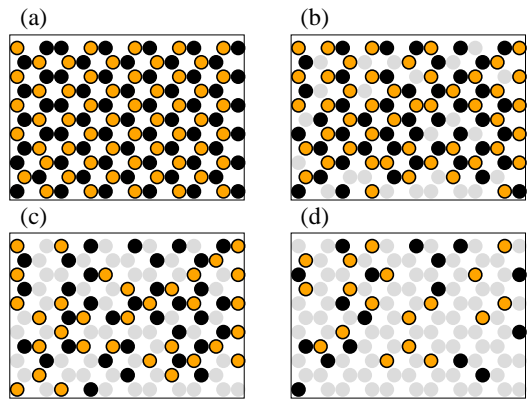}

\caption{Spin configuration on top of a hill for (a) homogeneous coverage and
(b) -(d) $h_{0}=$ -1, 0 and 0.28 $z_{max}$ with adatom concentrations
$x=0.5,\,0.11$ and 0.02 respectively. In black, spin up; in orange,
spin down. ($k_{B}T=0.01J_{0}$). \label{fig2}}

\end{figure}

\begin{figure}[t]
 \includegraphics[clip,width=0.75\columnwidth]{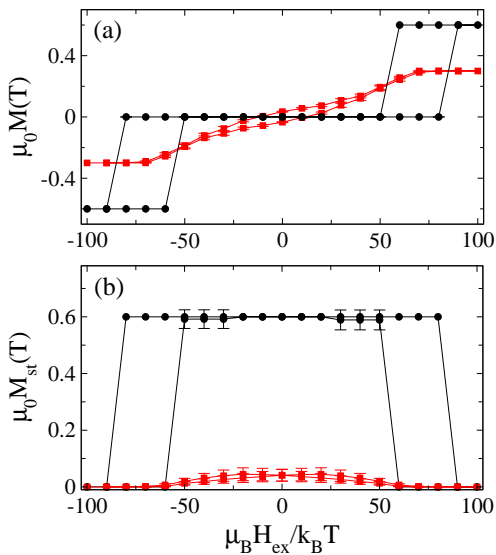}

\caption{Left: (a) Magnetization and (b) staggered magnetization as a function
of an external magnetic field for a fully covered graphene sheet (black
circles) and for $h_{0}$=0 (red squares) with $k_{B}T=0.01J_{0}$.
\label{fig3}}

\end{figure}

For a graphene sheet with full coverage, as shown in Fig.~\ref{fig2}
(a), we find that the moments order antiferromagnetically, as expected.
Nevertheless, when we associate the local probability of adsorption
of the adatoms to the ripples and we start to dilute the coverage
for increasing values of $h_{0}$, the situation is rather more complicated.
What we observe in figure \ref{fig2} (b)-(d) is the destruction of
the antiferromagnetic order on the clusters that are located on top
of the ripples. The system becomes highly \textit{frustrated} and
finally for very low coverage we find that the system consists basically
of weakly interacting isolated moments and clusters of few spins.
One notices that the destruction of antiferromagnetic order is accompanied
by a ferromagnetic tendency, that is, the clusters tend to \textit{either
remain antiferromagnetically ordered and interact ferromagnetically
with nearest clusters} (see Fig. \ref{fig2} (c) ) or order ferromagnetically
in a situation that is reminiscent either of \textit{super-paramagnetism}
or a \textit{spin glass} state \cite{mydosh}. \char`\"{}In this
paper we have considered only the average magnetic properties while
disregarding rare events that can lead to Griffiths phases \ref{griffiths}.
However, rare event physics can play an important role in these disordered
magnetic systems and may contribute to the magnetic response. This
is a problem, however, beyond the scope of the current manuscript.

For the configurations discussed above, we calculate the total average
magnetization $M=M_{A}+M_{B}$ and the average staggered magnetization,
$M_{s}=|M_{A}-M_{B}|$, in terms of the magnetization of each sublattice,
$M_{A}$ and $M_{B}$, as a function of the magnetic field $H_{ex}$.
All the numerical simulations were performed for an Ising Hamiltonian
with 2 $\times$ 600 $\times$ 600 sites. We used 10000 Monte Carlo
steps for warm up and another 10000 steps to collect the data. The
magnetization is computed as an average over 160 to 400 realizations.
As can be seen in Fig. \ref{fig3}(a) and (b) the maximally covered
honeycomb lattice has always an antiferromagnetic correlation between
the spins but as the coverage is reduced, this correlation is strongly
suppressed, changing the shape of the hysteresis loops. %
\begin{figure}[t]
\vspace{-0.3cm}
 \includegraphics[clip,width=0.8\columnwidth]{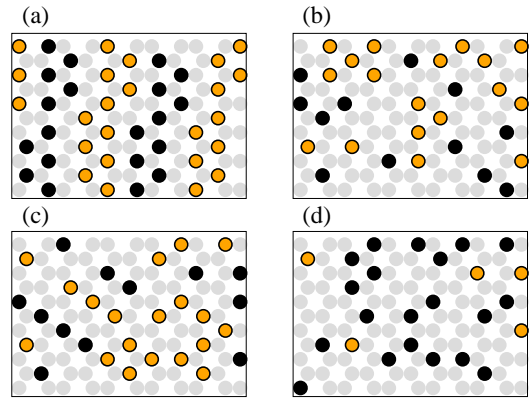}

\caption{Spin configurations for the same hill considered in figure \protect\ref{fig2}
but with the random removal of nearest neighbors. (a) homogeneous
honeycomb lattice, and (b) -(d) $h_{0}=-1,\,0$ and 0.28$z_{max}$
respectively. The adatom concentrations are $x=0.24$ in the homogeneous
lattice (a), $x=0.22$ (b), $x=0.08$ (c) and $x=0.01$ (d). In black,
spin up; in orange, spin down ($k_{B}T=0.01J_{0}$). \label{fig4}}

\end{figure}

\begin{figure}[b]
 \includegraphics[clip,width=0.85\columnwidth]{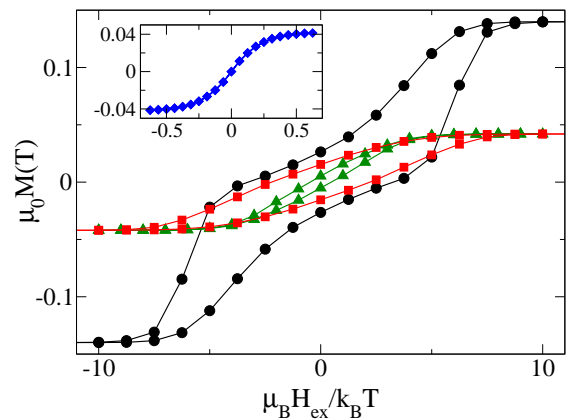}

\caption{ Magnetization as a function of magnetic field for a lattice without
first neighbors: homogeneous lattice (black circles), $h_{0}$=0 and
$k_{B}T=0.01J_{0}$ (red squares) and $k_{B}T=0.05J_{0}$ (green triangles).
Inset: $h_{0}$=0 and $k_{B}T=0.2J_{0}$ (blue diamonds). \label{fig5}}

\end{figure}

If we suppress the existence of nearest neighbors, one favors ferromagnetic
interactions, once the interaction between next-nearest neighbors
is ferromagnetic. However, as the magnetic atoms are randomly removed,
we never produce a regular lattice of spins and do not have a perfect
ferromagnetic system. Instead, for the system that was originally
a regular honeycomb lattice, we see in figure \ref{fig4} (a) the
formation of \textit{magnetic domains} in the form of antiferromagnetically
correlated \textit{ferromagnetic stripes}. In the dilute situation
{[}see Fig. \ref{fig4}(b)-(d)] these ferromagnetic stripes are broken
apart leaving behind ferromagnetic strings. The situation is clearly
quite complex from the magnetic point of view and it is important
to distinguish these various spin textures experimentally. %
\begin{figure*}[t]
 \vspace{-0.3cm}
 \includegraphics[clip,scale=0.5]{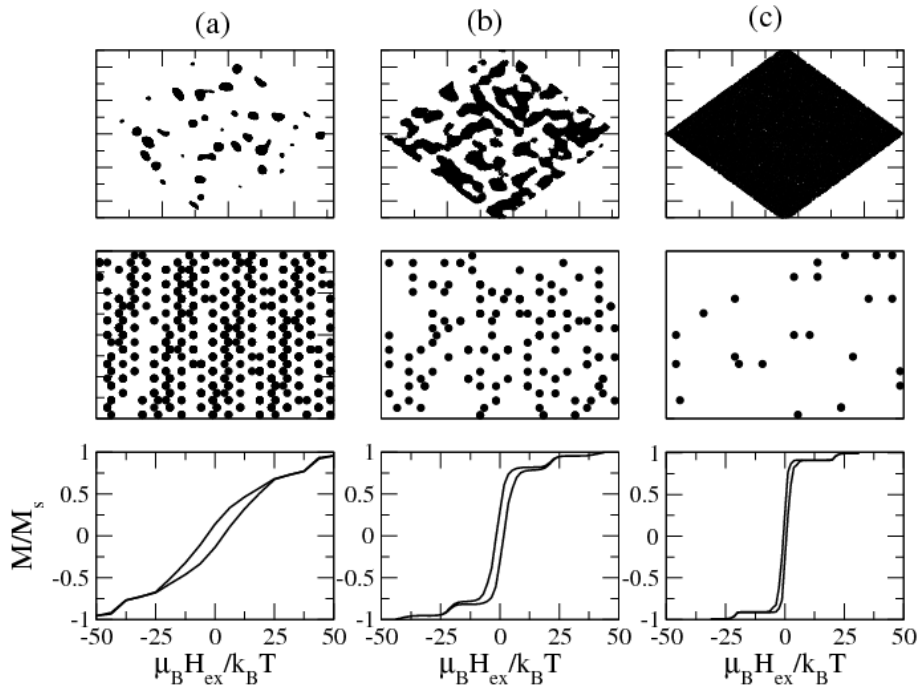}

\caption{ The three columns show the spin configuration (top), the details of the
configuration for a small region of the system (middle) and the total magnetization
versus magnetic field for (a) $h_{0}=0.28\, z_{max}$
(b), $h_{0}=0$ and (c) a random distribution. The three systems
contain the same concentration of adatoms, $x=0.05$. \label{fig6}}

\end{figure*}

In the case of an originally regular honeycomb lattice without neighboring
adatoms, the magnetization $M(H)$ in Fig.~\ref{fig5} shows a large
coercivity due to the ferromagnetic stripes (black circles), which
decreases with smaller incorporation fractions. Whereas for large
incorporation fractions we still can see the signature of antiferromagnetic
correlations in the shape of the hysteresis curve, for smaller incorporation
fractions the systems consists of ferromagnetic clusters with a typical
ferromagnetic hysteresis loop (green triangles and red squares). By
increasing temperature, we observe a decrease in the coercivity of
the hysteresis loop. For $k_{B}T\gtrsim0.2J_{0}$ (see inset of Fig.~\ref{fig5}),
the spins interact weakly and we obtain a hysteresis loop that resembles
a paramagnetic behavior. In this case, for low magnetic fields, $H_{ex}$,
we find a universal linear dependence in the magnetization, $M\propto H_{ex}$.

It is important to point out the differences between the type of magnetic
dilution we present here, compared to the usually magnetic diluted
lattice, where the spins are randomly located, without any preferential
position. In honeycomb lattices, randomly diluted spin systems have
a site percolation transition at $p_{c}=0.69704$~\cite{perco}.
For nearest neighbor ferromagnetic interactions, the occurrence of
a magnetic transition as a function of the dilution $p$ has been
shown to coincide with the percolation transition $p_{c}$~\cite{mcgurn}.
For the antiferromagnetic case, recent results for the quantum Heisenberg
Hamiltonian with site dilution~\cite{castro} found that the magnetic
long range order persists above $p_{c}$. In any case, in the randomly
diluted system the moments are always isolated and weakly coupled
at very small concentrations ($p\ll p_{c}$). In this work, we have
shown that due to the distribution of spins according to the structure
of the ripples, we suggest the possible existence of strongly coupled
magnetic clusters on top of the highest graphene hills, even at very
low concentration of adatoms. Also, due to the ferromagnetic coupling
between next nearest neighbors, the moment of a given cluster will
be bigger than zero even for a mostly antiferromagnetic one.

In Figure~\ref{fig6}, we compare the coverage structure and magnetization
curves for three systems with the same adatom concentration, $x=0.05$,
but different incorporation probability distributions. The total concentration
is the same in the three systems but $h_{0}$ sets regions where the
presence of adatoms is forbidden. As a result, we can see the presence
of strongly correlated clusters for $h_{0}=0.28\, z_{max}$, while
for a random distribution of adatoms we found almost isolated spins.
In addition, one can see the clear difference between the hysteresis
loop in the correlated case (ripples), shown in Fig. \ref{fig6} (a),
(b), compared to the completely random one {[}Fig. \ref{fig6} (c)].
In the correlated case, we found regions with high concentration of
spins and very strong antiferromagnetic correlations between them
at low temperatures ($k_{B}T\ll J_{0}$), in contrast with the random
case, where both hysteresis and the coercitive fields are very small.

\section{Magnetoresistance\label{sec4}}

The interplay between the corrugated nature of graphene and its magnetic
properties, which gives rise to the spin textures discussed above,
can be probed by magneto-transport measurements. It was recently demonstrated
that the incorporation of adatoms can dramatically change the transport
properties in graphene~\cite{elias09}, where the system can go
from a metallic behavior to a variable range hopping regime. In the
case of magnetic impurities, due to the exchange interaction between
the spin of the carriers and the adatom localized moments, the transport
properties will be further modified by the interaction between spin
and charge degrees of freedom. For strongly disordered graphene, which
characterizes the regime of variable range hopping, the carriers are
trapped with a binding energy $E_{i}$. Even at $H_{ex}=0$, the exchange
interaction fully polarizes the localized spins interacting with a
given localized carrier and increases its binding energy~\cite{dietl}.

If one considers the localized spins in the mean-field approximation,
the exchange energy is given by $E_{ex}=J_{k}M_{s}\langle s\rangle$
where $J_{k}$ is the exchange interaction between localized spins
and carriers, $M_{s}$ is the saturation magnetization of the magnetic
impurity spins and $\langle s\rangle$ is the mean value of the localized
carriers spin. The total binding energy in the presence of the exchange
interaction is given by $E_{b}=E_{i}+E_{ex}$ and has an associated
Bohr radius $\xi$. It this situation, each spin is fully polarized
in a random direction. If we apply an external field $H_{ex}$, the
localized spins in the whole sample begin to align with the field
and because of the exchange interaction between them and the band
carriers, $H_{ex}$ gives rise to a splitting of the conduction band.
As a consequence, there is a \textit{decrease} in the binding energy
by $\Delta E_{ex}=J_{k}M(T,H)\langle s\rangle$. As $E_{b}\propto\xi^{-2}$,
the magnetic field produces an increase in the effective Bohr radius
\begin{equation}
\xi_{{\rm eff}}(T,H)=(E_{i}+J_{k}\langle s\rangle M_{s}[1-|M(T,H)|/M_{s}])^{-1/2}\,.\end{equation}
 This effect is similar to the one observed in magnetically semiconductors
as EuS or CdMnTe~\cite{dietl}.

\begin{figure}[b]

\begin{centering}
\includegraphics[width=0.9\columnwidth]{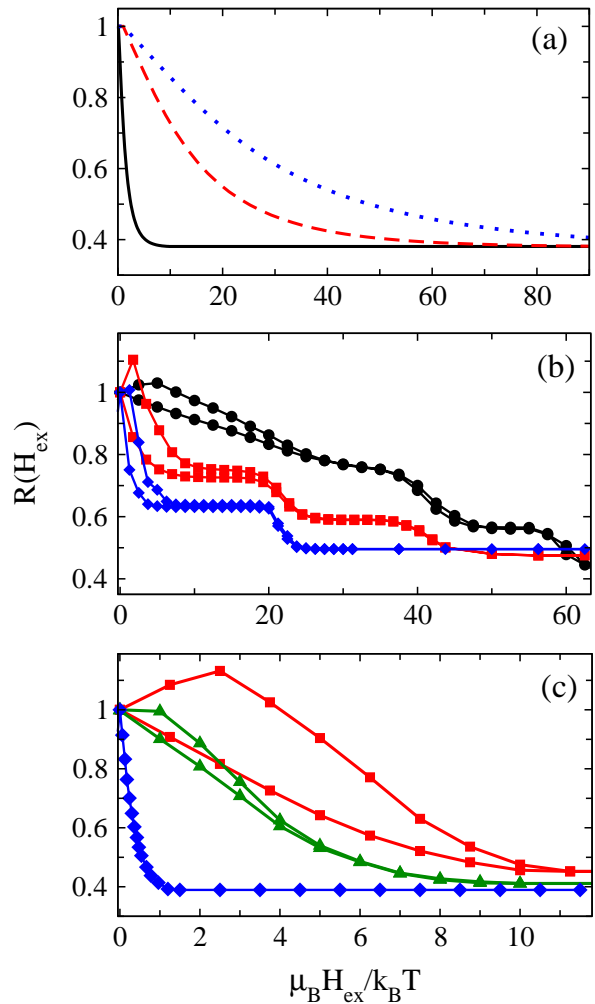} 
\par\end{centering}

\vspace{0.1cm}

\caption{(a) Magnetoresistance for non-interacting spins at three different
temperatures: $k_{B}T=0.01J_{0}$(solid line), $k_{B}T=0.1J_{0}$
(dashed line) and $k_{B}T=0.2J_{0}$ (dotted line). (b) Magnetoresistance
for the three different incorporation distributions shown in Fig~\protect{\ref{fig6}}:
$h_{0}$=0.28 $z_{max}$ (black circles), $h_{0}$=0 (red squares)
and a random distribution (blue diamonds). (c) Magnetoresistance
in the case without first neighbor H atoms for $h_{0}$=0 and different
temperatures: $k_{B}T=0.01J_{0}$ (red squares), $k_{B}T=0.05J_{0}$
(green triangles) and $k_{B}T=0.2J_{0}$ (blue diamonds). All curves
are normalized by the corresponding zero field resistance $R(H_{ex}=0)=1$
and are symmetric for $H_{ex}<0$.\label{fig7} }

\end{figure}

The hopping probability between two states at a distance $r$ is then
given by $P=\exp(-2r/\xi_{{\rm eff}}-W/k_{B}T$) where $W$ is the
energy difference between the two states and $\xi_{{\rm eff}}$ is
the characteristic Bohr radius of the localized states under the effect
of an external magnetic field. Following the original Mott derivation,
the resistance is\cite{mott}, \begin{equation}
\rho=\rho_{0}\exp(T_{0}(H)/T)^{1/3}\,,\label{rho}\end{equation}
 where \begin{equation}
T_{0}(H)\propto13.8/\left(k_{B}N(\mu)\xi_{{\rm eff}}^{2}\right),\end{equation}
 and $N(\mu)$ is the density of states at the Fermi energy. Since
$T_{0}\propto1-|M(T,H)|/M_{s}$ for small magnetization, we can extract
the magnetoresistance curves of the system from our previous Monte
Carlo results. In all curves, we normalize the scale of the magnetoresistance
by the zero field case, $R(H_{ex}=0)=1$. From the experimental point
of view, the change in the temperature and gate voltage parameters
in the range where Eq. (\ref{rho}) is applicable will additionally
rescale the magnetoresistance curves. The scaling analysis of the
magnetoresistance provides additional information allowing the empirical
determination of the exchange interaction $J_{k}$ and the binding
energy $E_{i}$.

We begin our analysis by considering non-interacting spins. In this
case, the response is paramagnetic and the magnetization curves follow
a typical Brillouin function \begin{equation}
M(H_{ex},T)\propto\frac{2S+1}{2S}\coth\!\left(\frac{2S+1}{2S}x\right)-\frac{1}{2S}\coth\!\left(\frac{1}{2S}x\right)\label{eq:M}\end{equation}
 where $x=g\mu_{B}SH_{ex}/k_{B}T$, and $S$ is the spin quantum number.
Assuming $g=2$ and $S=1/2$, we show in Fig~\ref{fig7}(a) the magnetoresistance
curves for a system with non-interacting spins at different temperatures.
The magnetoresistance is negative, reaching a minimum value when all
the spins are aligned by the magnetic field. For a fixed magnetic
field, the resistance grows with temperature.

Next, we present the magnetoresistance curves corresponding to the
simulations in the previous sections. If we allow the presence of
neighboring magnetic adatoms, the antiferromagnetic correlations between
neighbors is strong and the minimum value of the magnetoresistance
is only obtained for very large values of the magnetic field, set
in energy units of $k_{B}T$. In Fig.~\ref{fig7}(b) we compare the
magnetoresistance curves that correspond to the configurations shown
in Fig.~\ref{fig6} for a fixed concentration of adatoms, $x=0.05$,
and different incorporation probabilities. In the correlated case,
{[}Fig.~\ref{fig6}(a),(b)], because of the fixed concentration of
adatoms, the top of the highest ripples accumulates larger antiferromagnetic
clusters for larger values of $h_{0}$, requiring the application
of stronger magnetic fields to decrease the resistivity. On the other
hand, for the case of random distribution of spins, the antiferromagnetic
correlations are much weaker. Notice that for systems with hysteretic
behavior in the magnetization, $M$ is zero at the coercive field.
Consequently, the magnetoresistance at the coercive field is higher
than its value for $H_{ex}=0$, giving rise to a \emph{positive} magnetoresistance
at small fields.

In the scenario without nearest H neighbors, we consider the probability
distribution with $h_{0}=0$, for different temperatures {[}see Fig~\ref{fig7}(c)].
For $k_{B}T\gtrsim0.2J_{0}$ the hysteretic nature of the magnetoresistance
is considerably reduced and we obtain a curve that is similar to the
case of non-interacting spins. Neverthless, we note that the width
of the resistivity peak is anomalously small, indicating that the
effective size of such``non-interacting'' moments is considerably
larger than the moment of isolated adatoms. This is a clear indication
for the presence of small magnetic clusters in graphene. For even
higher temperatures, $k_{B}T\sim J_{0}$, the resistance decreases
with $H$ following $\rho\propto\exp(-\alpha H)$, where $\alpha$
is a constant that depends on the exchange interaction, density of
states and temperature. For weak external fields $H_{ex}\ll\Delta_{SO}$,
where $\Delta_{SO}$ is the spin orbit coupling, the magnetoresistance
will be strongly anisotropic and can vanish when the magnetic field
is parallel to the graphene plane.

\section{Conclusions}

In summary, we have discussed the possible magnetic states of magnetic
adatoms on top of a rippled graphene sheet. We proposed the scenario
where the non-trivial topography of graphene correlates the local
moments of the adatoms, generating clusters and non-trivial macroscopic
magnetic states. The magnetic order can be very sensitive to the value
of the chemical potential, the adatom coverage, and the {}``repulsion''
between adatoms for nearest neighbor atoms.

We have found that while a perfectly covered graphene sheet has a
strong tendency towards antiferromagnetism, the presence of ripples
leads to more complex magnetic textures, with increased ferromagnetic
tendency and even glassiness. We show that the hysteretic behavior
of the magnetization provides a simple way to study these magnetic
orderings. Nevertheless, the presence of different magnetic structures
affect directly the transport properties. In the variable range hopping
regime, the system presents a universal negative magnetoresistance
that depending on the ratio between the exchange interaction, temperature
and coverage fraction can have also show hysteretic behavior. The
magnetotransport curves in graphene can provide clear experimental
signatures for these non-trivial magnetic states.

We thank A. Geim, K. Novoselov, E. Fradkin and M. B. Silva Neto for
many level headed discussions. AHCN acknowledges the partial support
of the U.S. Department of Energy under the grant DE-FG02-08ER46512.
BU acknowledges partial support from Office of Science, U.S. Department
of Energy under the grant DE-FG02-91ER45439 at University of Illinois.
TGR acknowledges the partial support of the brazilian agencies CNPq
and FAPERJ and L'Oreal Brazil.

\end{document}